# Mapping the functional connectome traits of levels of consciousness.

Enrico Amico[a,b], Daniele Marinazzo[b], Carol Di Perri[a,c], Lizette Heine[a,c], Jitka Annen[a,c], Charlotte Martial[a,c], Mario Dzemidzic[d], Murielle Kirsch[c], Vincent Bonhomme[c], Steven Laureys[a,c,*] and Joaquín Goñi[e,f,g,*]

[a]Coma Science Group, GIGA Research Center, University of Liège, Liège, Belgium

[b]Department of Data-analysis, University of Ghent, B9000 Ghent, Belgium

[c]University Hospital of Liège, Liège, Belgium

[d]Department of Neurology and Radiology and Imaging Sciences, Indiana University School of Medicine, Indianapolis, IN, USA

[e]School of Industrial Engineering, Purdue University, West-Lafayette, IN, USA

[f]Weldon School of Biomedical Engineering, Purdue University, West-Lafayette, IN, USA

[g]Purdue Institute for Integrative Neuroscience, Purdue University, West-Lafayette, IN, USA

*Authors contributed equally.

steven.laureys@ulg.ac.be

jgonicor@purdue.edu

# Abstract

Examining task-free functional connectivity (FC) in the human brain offers insights on how spontaneous integration and segregation of information relate to human cognition, and how this organization may be altered in different conditions, and neurological disorders. This is particularly relevant for patients in disorders of consciousness (DOC) following severe acquired brain damage and coma, one of the most devastating conditions in modern medical care.

We present a novel data-driven methodology, *connICA*, which implements Independent Component Analysis (ICA) for the extraction of robust independent FC patterns (FC-traits) from a set of individual functional connectomes, without imposing any a priori data stratification into groups.

We here apply *connICA* to investigate associations between network traits derived from task-free FC and cognitive/clinical features that define levels of consciousness. Three main independent FC-traits were identified and linked to consciousness-related clinical features. The first one represents the functional configuration of a "resting" human brain, and it is associated to a sedative (sevoflurane), the overall effect of the pathology and the level of arousal. The second FC-trait reflects the disconnection of the visual and sensory-motor connectivity patterns. It also relates to the time since the insult and to the ability of communicating with the external environment. The third FC-trait isolates the connectivity pattern encompassing the fronto-parietal and the default-mode network areas as well as the interaction between left and right hemispheres, which are also associated to the awareness of the self and its surroundings.

Each FC-trait represents a distinct functional process with a role in the degradation of conscious states of functional brain networks, shedding further light on the functional sub-circuits that get disrupted in severe brain-damage.

## Introduction

Disorders of consciousness (DOC) remain among the most challenging and poorly understood conditions in modern medical care. The term spreads over several pathological states qualified by dissociation between awareness and arousal (Bernat, 2009; Laureys, 2005). Among these, patients in coma show no signs of awareness nor arousal; patients with unresponsive wakefulness syndrome/vegetative state (UWS) show no signs of awareness but do have an altered sleep and wake cycle; patients in a minimally conscious state (MCS) retain minimal non-reflexive and highly fluctuating signs of awareness. When patients regain functional object use and/or reliable communication they are referred to as emerging from MCS (EMCS) (Giacino et al., 2014; Laureys et al., 2004). A particular outcome is represented by patients with a locked-in syndrome (LIS), who have no means of producing speech, limb or facial movements (except mostly for eye movement and/or blinking) but are still awake and fully conscious (Giacino et al., 1995; Laureys et al., 2005). To date, the most validated diagnosis of these patients is based on the behavioral presentation of the patient. The distinction between these pathological levels of consciousness can be very challenging, as the boundaries between these states are often uncertain and ambiguous (Giacino et al., 2014).

In the last decade, advances in neuroimaging techniques have allowed the medical community to gain important insights into the pathophysiology of DOC and to observe that altered states of consciousness are related to complex disruptions in the functioñal and structural organization of the brain (Boly et al., 2012; Di Perri et al., 2014; Fernández Espejo et al., 2012; Koch et al., 2016; Owen et al., 2009).

At the same time, quantitative analysis based on complex networks have become more commonly used to study the brain as a network (Bullmore and Sporns, 2009), giving rise to the area of research so called Brain Connectomics (Fornito et al., 2016; Sporns, 2011). In brain network models, nodes correspond to grey-matter regions (based on brain atlases or parcellations) while links or edges correspond to connections. Structural connections are modeled using white matter fiber-tracts and functional connections represent coupling between brain regions while subjects are either at rest or performing a task (van den Heuvel and Hulshoff Pol, 2010). Recent advances in functional neuroimaging have provided new tools to measure and examine *in vivo* whole-brain temporal dependence of the dynamics of anatomically separated brain regions, defined as functional connectivity (FC) (Fox and Raichle, 2007; Fox et al., 2005; Friston et al., 1993).

In parallel to the development of methods and network features in Brain Connectomics, analyses of functional magnetic resonance imaging (fMRI) data based on independent component analysis (ICA) have become an increasingly popular voxel-level approach (Calhoun et al., 2009). ICA, by relying upon a general assumption of the independence of the mixed signals, is a powerful and versatile data-driven approach for studying the brain, at both temporal and spatial scales (Erhardt et al., 2011).

Examining functional connectivity in the human brain offers unique insights on how integration and segregation of information relates to human behavior and how this organization may be altered in diseases (Boly et al., 2012; Greicius, 2008). In the case of disorders of consciousness, voxel-level ICA-based fMRI studies of levels of consciousness in DOC patients have mainly shown alterations in the functional connectivity of the default mode network (DMN) (Heine et al., 2012; Soddu et al., 2012; Vanhaudenhuyse et al., 2010). Recent studies have also shown disrupted functional connectivity in resting state networks other than DMN (Demertzi et al., 2014) and possibility to correctly classify patients based on the level of connectivity of the "auditory" network (Demertzi et al., 2015). Furthermore, analyses of the functional networks of comatose brains have also evidenced a radical reorganization of high degree "hub" regions (Achard et al., 2012) and also showing that most of the affected regions in patients belonged to highly interconnected central nodes (Crone et al., 2014; Koch et al., 2016).

The potential of functional connectivity (FC) in particular and of Brain Connectomics in general in exploring the diseased human brain as a network going through systemic changes is undisputed. However, there is still no clear way to accomplish two critical steps of great clinical importance. First, to separate underlying FC patterns representing different functional mechanisms and, second, to relate those FC patterns or subsequent network features to individual cognitive performance or clinical evaluations. This is specially the case when studying a continuum of states, where the stratification of the cohort-subjects into categories or groups is inappropriate and/or poorly defined. Furthermore, standard FC techniques are not able to model and disentangle common underlying forces or competing processes arising from different functional patterns of healthy and diseased human brains in a data-driven fashion, as for instance ICA does in the case of fMRI voxel time series (Calhoun et al., 2009; Erhardt et al., 2011). This was indeed our motivation for the approach presented here.

In this study we bridge this gap by presenting a novel data-driven methodology, *connICA*, which consists of the extraction of robust independent patterns (traits) from a set of individual functional connectomes (see scheme in Figure 1). In this sense, *connICA* is a multiplex network framework both in the input (i.e., layers are individual FC connectomes) and in the output (i.e., layers are independent patterns or FC-traits). Here we apply *connICA* to investigate the link between cognitive/clinical features that define states of consciousness and resting-state functional connectivity (FC) data. This method allows the assessment of individual FC patterns (or FC layers) in a joint data-driven fashion providing as outputs multivariate independent FC-traits, which model independent sources or phenomena present in the input (i.e. the aforementioned individual FC patterns). In a final step, we assess the predictability of the weights (fingerprints) of each FC-trait on each subject from demographic and consciousness related variables, allowing for a continuous mapping of levels of consciousness within functional connectomes.

## Materials and methods

### Subjects

The cohort studied here consists of 88 subjects with different levels of consciousness. From those, 31 were healthy controls (mean age 44 years ± 15 years, 20 males, 11 females). We included 57 patients from an initial cohort of 216 patients in different levels of consciousness. Exclusion criteria were: 1) neuroimaging examination in an acute state, i.e. <28 days from brain insult, 2) large focal brain damage, i.e. >2/3 of one hemisphere, as stated by a certified neuroradiologist, 3) suboptimal segmentation, normalization and/or parcellation of the brain volumes after visual inspection. Out of the selected 57, 39 were patients with disorders of consciousness (2 coma, 17 UWS, 21 MCS), 13 EMCS and 4 LIS. Also, 28 out of 57 patients had traumatic brain injury (TBI), and 30 were sedated during the fMRI acquisition (please see Demographics section for details).

Healthy volunteers were free of psychiatric or neurological history. The study was approved by the Ethics Committee of the Medical School of the University of Liège. Written informed consent to participate in the study was obtained from the healthy subjects and from the legal surrogates of the patients.

**Demographics**

Nuisance variables included age, gender, etiology (1 for TBI, 0 otherwise), sedation and the inverse of the time (in days) since onset (i.e. insult), as we assumed healthy subjects' time since onset to be infinite and hence corresponding to zero in our codification.

The presence of sedated and non-sedated patients in the sample has to be taken into account as a major confound, since the "sedation" effect in resting state FC has been shown to depend on the specific sedative, including differences between sedation and natural sleep (Fiset et al., 2005; Laureys, 2005; Palanca et al., 2015; Tagliazucchi et al., 2012). Furthermore, recent studies have established that the depth of sedation has an effect on FC measured using task-free fMRI (Monti et al., 2013; Stamatakis et al., 2010).

In the study presented here, patients were scanned on clinical demand, and the magnetic resonance imaging (MRI) acquisition was performed without sedation whenever possible. In case of excessive movements of the patient in the scanner, the senior anesthesiologist determined, based on the clinical history, to administer light sedation to the patient by using the minimum necessary dose. The type of sedative chosen by the senior anesthesiologist, for the sedated patients included in this cohort, was either propofol (N=18) or sevoflurane (N=9), or the combination of the two (N=3). Propofol was always administered intravenously (1-2 μg/mL), using a target-controlled infusion system allowing targeting a precise plasma concentration, and based on a pharmacokinetic model. Sevoflurane was given through inhalation (1-2% concentration), in spontaneously breathing subjects. The choice of the sedative agent was left at the discretion of the anesthesiologist in charge, taking into account the presence of devices for controlling the airway (endotracheal tube or tracheostomy) and allowing the easy administration of inhaled agents. Since the concentrations of the sedatives were comparable across subjects, we decided to factor the confound effects as two independent binary variables, one for sevoflurane (sevof), and one for propofol (propo).

To assess the level of consciousness, we used the scores obtained from the JFK Coma Recovery Scale-Revised (CRS-R) (Giacino and Kalmar, 2006; Kalmar and Giacino, 2005; Schnakers et al., 2008) assessment for each DOC patient. The CRS-R is the most sensitive and validated (Seel et al., 2010) scale to fully characterize and monitor DOC patients and provide a global quantification of their levels of consciousness. In particular, CRS-R integrates 25 arranged items that comprise 6 sub-scales addressing auditory, visual, motor, oromotor, communication, and arousal processes. Each item assesses the presence or absence of a specific physical sign that represents the integrity of brain function at one of four levels: generalized, localized, emergent, or cognitively mediated responsiveness. Scoring is based on the presence or absence of specific behavioral responses to sensory stimuli administered in a standardized manner. The reader can refer to (Giacino and Kalmar, 2006; Giacino et al., 1991; Schnakers et al., 2008) for a detailed description of the scale.

**Image acquisition**

Each subject underwent structural MRI and a 10 minute fMRI resting-state (task-free) session. Whole-brain structural MRI T1 data (T1-weighted 3D MP-RAGE, 120 transversal slices, repetition time = 2300 ms, voxel size = 1.0 x 1.0 x 1.2 mm$^3$, flip angle = 9°, field of view = 256 x 256 mm$^2$ ) and resting state Blood-oxygenation-level dependent (BOLD) fMRI data (Echo Planar Imaging sequence, gradient echo, volumes = 300, repetition time = 2000 ms, echo time = 30 ms, flip angle = 78°, voxel size = 3 x 3 x 3 mm$^3$, field of view = 192×192 mm$^2$, 32 transversal slices) were acquired on a Siemens 3T Trio scanner. Healthy subjects were instructed to keep eyes open during the fMRI acquisition.

**Data processing and Functional Connectivity modeling**

Data processing was performed by combining functions from FSL (Jenkinson et al., 2012) and in-house developed Matlab (MATLAB 6.1, The MathWorks Inc., Natick, MA, 2000) code. The individual functional connectomes were modeled in the native BOLD fMRI space of each subject.

Processing steps were based on state of the art fMRI processing guidelines (Power et al., 2012; Power et al., 2014). Structural images were first denoised to improve the signal-to-noise ratio (Coupé et al., 2012), bias-field corrected, and then segmented (FSL FAST) to extract white matter, grey matter and cerebrospinal fluid (CSF) tissue masks. These masks were warped in each individual subject's functional space by means of subsequent linear and non-linear registrations (FSL flirt 6dof, FSL flirt 12dof and fnirt).

BOLD fMRI functional volumes were processed according to the steps recommended by (Power et al., 2014). These steps included: slice timing correction, motion correction, normalization to mode 1000, demeaning and linear detrending, inclusion of 18 regressors consisting of 3 translations [x,y,z], 3 rotations [pitch, yaw, roll], and 3 tissue regressors (mean signal of whole-brain, white matter (WM) and cerebrospinal fluid (CSF)), and the 9 corresponding derivatives (backwards difference, see Figure S1). A scrubbing procedure censoring high head motion volumes was based on Frame Displacement (FD), DVARS and SD metrics. FD measures the movement of the head from one volume to the next, and is calculated as the sum of the absolute values of the differentiated realignment estimates (by backwards differences) at every time-point (Power et al., 2014); DVARS (D referring to temporal derivative of BOLD time courses, VARS referring to root mean square variance over voxels) measures the change in signal intensity from one volume to the next, and is calculated as the root mean square value of the differentiated BOLD time-series (by backwards differences) within a spatial mask at every time-point (Smyser et al., 2011); SD stands for the standard deviation of the BOLD signal within brain voxels at every time-point (outlier volumes higher than 75 percentile +1.5 of the interquartile range were discarded, see Fig. S1). There were no significant differences in the number of volumes censored between controls and patients (Wilcoxon ranksum test, p=0.18).

A bandpass first-order Butterworth filter in forward and reverse directions [0.001 Hz, 0.08 Hz] was then applied. After that, the 3 principal components of the BOLD signal in the WM and CSF tissue were regressed out of the gray matter (GM) signal.

A whole-brain data-driven functional parcellation based on 278 regions, as obtained by Shen and colleagues (Shen et al., 2013), was first warped into each subject's T1 space (FSL flirt 6dof, FSL flirt 12dof and finally FSL fnirt) and then into each subject's fMRI space. To improve the registration of the structural masks and the parcellation to the functional volumes FSL boundary-based-registration (Greve and Fischl, 2009) was also applied. Individual functional connectivity matrices (FC) were then estimated by means of pairwise Pearson correlations between the averaged signals of the regions of the parcellation, excluding the censored volumes as determined by the above-mentioned scrubbing procedure. We did not perform voxel-level spatial smoothing prior to averaging of the voxel time-series per region because spatial smoothing with a Full-Width Half-Maximum of 4mm isotropic Gaussian kernel produced almost unnoticeable differences (results not shown).

Finally, the resulting individual FC matrices were ordered according to 7 resting-state cortical sub-networks (*RSN*s) as proposed by Yeo and colleagues (Yeo et al., 2011) (see insert of Fig 2B). For completeness, we added two more sub-networks: an $8^{th}$ sub-network comprised of the subcortical regions and a $9^{th}$ sub-network including the cerebellar regions.

### *ConnICA*: Independent component analyses of sets of individual functional connectomes

The input of *ConnICA* consists of all the individual FC profiles embedded into a dataset matrix where each row contains all the entries of the upper triangular part of the FC matrix for each subject (given the symmetry of FC) and hence provides an individual FC pattern. Note that this includes all FC matrices from all subjects, without any *a priori* information or any stratification of the data into groups (see scheme at Fig. 1). With this input, ICA decomposition of the FC patterns was applied by running FastICA algorithm (Hyvarinen, 1999) and setting the number of independent components to 15.

The output of *connICA* consists of two vectors per component. The first output vector will be referred to as *FC-trait*, which represents an independent pattern of functional connectivity. Interestingly, this vector can be represented back to its *spatial* form, i.e. a square symmetric matrix with brain regions in rows and columns. While the values here express connectivity units, they are not Pearson correlation coefficients and hence not restricted to the [-1,1] range. The second output vector is the weight of the FC-trait on each subject, which quantifies the prominence or presence of the trait in each individual FC matrix (note that this value can be positive or negative). In that sense, *connICA* is maximizing the individual variance explained by the multi-linear regression of the obtained ensemble of FC-traits and subsequent subject weights.

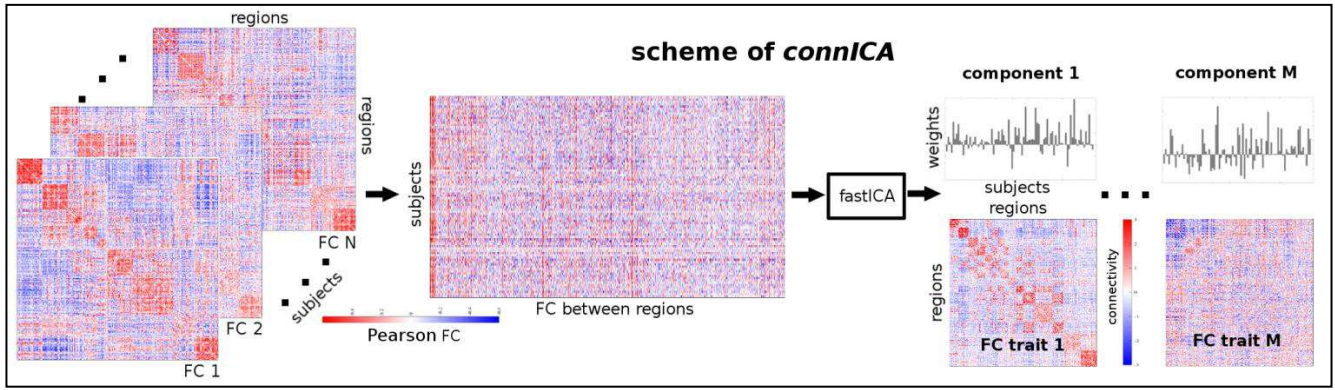

**Figure 1. Workflow scheme of the Connectivity Independent Component Analysis (*connICA*).** The upper triangular of each individual functional connectivity (FC) matrix (left) is added to a matrix where rows are the subjects and columns are their vectorized functional connectivity patterns. The ICA algorithm extracts the M independent components (i.e. functional traits) associated to the whole population and their relative weights across subjects. Colorbars indicate positive (red) and negative (blue) connectivity values, being Pearson's correlation coefficient values in the case of individual FC matrices (left side of scheme), and unitless connectivity weights in the case of FC-traits (right side of the scheme).

Given the non-deterministic nature of the FastICA decomposition into components (Hyvarinen, 1999), it was very important to run it several times and only select the most robust ones (from now on simply denominated FC-traits).Therefore, rather than analyzing the *connICA* components from a single FastICA run, we evaluated their similarity over 100 runs. For an FC-trait to be robust, it has to appear (correlation of 0.75 or higher across runs) in at least 75% of the runs. This procedure was divided in two steps: first, traits from each runs were compared all-to-all so to identify traits "similar" to each other across runs (i.e., spatial correlation ≥ 0.75); second, the frequency of these similar traits was computed and the ones that did not appear in at least 75% of the runs were discarded. The similar single-run traits that survived to this threshold (i.e., the most frequent ones), were then averaged together, in order to obtain mean robust traits across all runs. This criterion resulted in 5 robust FC-traits. (see Figure S2).

Each robust FC-trait was characterized by the mean and standard deviation of explained variance with respect to the individual FC matrices. The subject weights associated to each assessed FC-traits were then used as response in an incremental multi-linear regression model. Predictors included the Coma Recovery Scale Revised (CRS-R) (30) clinical subscores of each patient (Arousal, Auditory, Communication, Motor, Oromotor, Visual), and the sum of these scores. The control population was assigned with the highest scores for each of the coma recovery subscales. As aforementioned, the following variables were also included: age, gender, etiology (traumatic/non traumatic), sedation level (a binary variable for sevoflurane, a binary variable for propofol, see Demographics section for details) and the inverse of the time (days) since onset. In order to increase the validity of our binary codification of both sedatives and, overall, be able to better account for sedation effects, six patients with higher or different dose regimes were excluded from the multi-linear modeling.

We then identified the FC-traits whose presence (weights) in individual FCs was significantly explained by a cognitive predictor (statistical significance set at p-value ≤ 0.05, Fig. 3G, 3H, 3I). The aim was to extract the connectivity patterns or traits associated to consciousness-related clinical features.

**Modularity analyses**

Modularity is a measure of the strength of division of a network into modules or communities. Networks with high modularity have dense connections between the nodes within modules but sparse connections between nodes in different modules. One of the network modularity metrics is the Newman-Girvan quality function Q, defined as the fraction of edges that fall within modules minus the expected number of edges for a random graph with the same node degree distribution as the given network (Newman and Girvan, 2004). Particularly, we here use the extension of Q for signed undirected networks proposed by Mucha et al. (Mucha et al., 2010), and inspired by others (Gómez et al., 2009; Traag and Bruggeman, 2009).

To investigate the functional organization properties of the FC-traits extracted with *connICA*, we first identified to what extent the organization into communities (*RSN*s) presented by Yeo et al. is reflected in each FC-trait. We first assessed the similarity of each trait with Yeo's partitions. To do so, we used Newman-Girvan modularity function Q for signed undirected networks (Mucha et al., 2010) as a *fitness* function of the RSNs partition into each FC-trait.

We then assessed the community structure of each FC-trait by using the Louvain method for identifying communities in large networks (Blondel et al., 2008). In order to improve the stability of the community detection procedure, we performed consensus clustering (Lancichinetti and Fortunato, 2012) out of a set of 100 partitions obtained by the Louvain method. The consensus clustering technique performs a search for a consensus partition, i.e. the partition that is most similar, on average, to all the input partitions. The similarity can be measured in several ways, for instance co-occurrence of the nodes in the clusters of the input partitions (Lancichinetti and Fortunato, 2012). This "consensus" partition was finally selected for being the most robust one.

## Results

Following individual subject BOLD fMRI data processing (see Figure S1 for examples of four individual sessions) and subsequent modeling of the individual task-free functional connectomes, *connICA* (see scheme at Figure 1) was applied to the cohort of 88 subjects (31 conscious controls and 57 severely brain-damaged patients at different levels of consciousness; see Methods for details) without imposing any *a priori* information or stratification into groups. The procedure including 100 runs of *connICA* generated five robust FC-traits present with high frequency and reproducibility across runs (see Figure S2 and methods for details). Each FC-trait consists of two elements: 1) an FC map of the unitless connectivity weights with the same dimensions as an individual FC matrix, and 2) a vector indicating the *amount* of the FC-trait present on each individual functional connectome (i.e. the weight of the trait on each subject). Importantly, this latter *connICA* outcome allows us to associate individual cognitive and clinical features to each trait. Each of these 5 components (FC-traits) was then evaluated in terms of explained FC variance and Newman's modularity quality function Q (Newman and Girvan, 2004) generalized for signed networks (Mucha et al., 2010) with respect to the partition into *RSN*s proposed by Yeo and colleagues (Yeo et al., 2011). The highest explained variance components were 1, 2 and 4, where a dominant FC-trait 1 explained 18% of variance on average (Figure 2). It is important to note that the explained variance of a given FC-trait does not imply its' meaningfulness with respect to the variables of interest (i.e. those related to levels of consciousness in this case), but only the average prominence of that trait in the set of the FC connectivity matrices extracted from the population of subjects.

Of the 5 extracted traits, both FC-traits 1 and 2 had a high modularity ratio Q score (see Figure 2A), which denotes their strong fingerprint on the underlying *RSN*s organization (see insert in Figure 2B) in functional communities. In subsequent analysis we focused on FC traits

1, 2 and 4, which had highest $R^2$ and at the same time captured different aspects of the *RSN*s modular architecture.

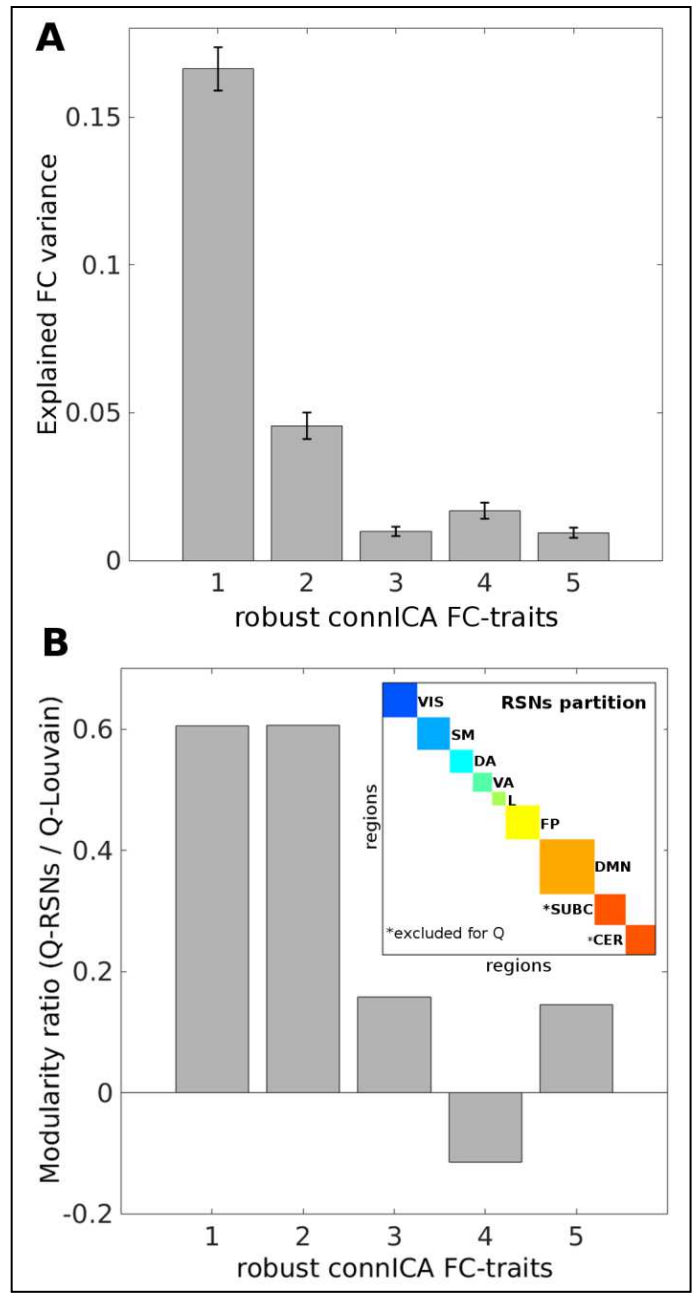

**Figure 2. *connICA*-extracted robust FC traits A) Bar-plot of the explained variance for the 5 most robust FC traits extracted with *connICA*.** Error bars show the standard error across subjects. **B) Bar-plot of the modularity ratio for the 5 robust FC-traits extracted with *connICA*.** This ratio is defined as the quality function Q ((Mucha et al., 2010), see Materials and Methods) for the imposed a priori Resting State Networks' (*RSN*s) partition (encompassing 7 networks: visual (VIS), sensorimotor (SM), dorsal attention (DA), ventral attention (VA), limbic (L), fronto-parietal (FP), default mode network (DMN), see top right insert in panel B), divided by the quality function Q (Mucha et al., 2010) for the data-driven partition obtained from consensus clustering (Lancichinetti and Fortunato, 2012) and Louvain algorithm (Blondel et al., 2008).

Given the novelty of our approach, we also tested: 1) whether the choice of the number of components (here set to 15) would "drive" the computation of the extracted FC-traits; 2) whether any subset of subjects would "drive" the computation of these traits; 3) whether the sedated patients would "drive" the computation of the extracted FC-traits. Overall, the results of these tests confirmed the robustness of the presence of the three FC-traits to the number of components, to data subsampling (randomly choosing 80% of subjects at a time) and to the exclusion of sedated patients. Please see Figure S3 of the Supplementary information for details.

The dominant FC-trait extracted using *connICA* (i.e. the one with the highest explained FC variance in the cohort) is shown in Figure 3D. Interestingly, it conforms to all the connectivity blocks or modules of the resting-state functional networks (*RSN*s, see insert Figure 2B) as introduced by Yeo and colleagues (Yeo et al., 2011). For this reason, this FC-trait was denominated the *RSNs* trait. We implemented an incremental multi-linear model predicting the weight or *quantity* of the *RSNs* trait on each subject (see Figure 3A) with up to 8 predictors (see Figure 3G).

We observed a significant negative association of the FC individual weights and the presence of sevoflurane sedative. That is, patients sedated with sevoflurane have a lower presence of the *RSNs* trait on their individual FC patterns. In addition, we found a significant positive association with the Coma Recovery Scale-Revised (CRS-R) (Giacino and Kalmar, 2006; Kalmar and Giacino, 2005; Schnakers et al., 2008) sum of scores while association with the arousal subscore was negative. That is, higher presence of the *RSNs* trait in subjects' functional connectome corresponded with higher CRS-R sum of scores and lower arousal subscore in those subjects.

The other two FC-traits linked to cognitive features associated with levels of consciousness (i.e. the communication subscore (Giacino and Kalmar, 2006; Kalmar and Giacino, 2005)) are shown in Figure 3E and 3F.

In particular, the FC-trait depicted in Figure 3E mainly captures changes of intra-hemispheric functional connectivity in the visual and sensory motor networks across subjects at different levels of consciousness. We will refer to it as the *VIS-SM* trait. A significant relationship with the CRS-R communication subscore (Giacino and Kalmar, 2006; Kalmar and Giacino, 2005) was found, as well as with the inverse of the time since onset (see Figure 3H).

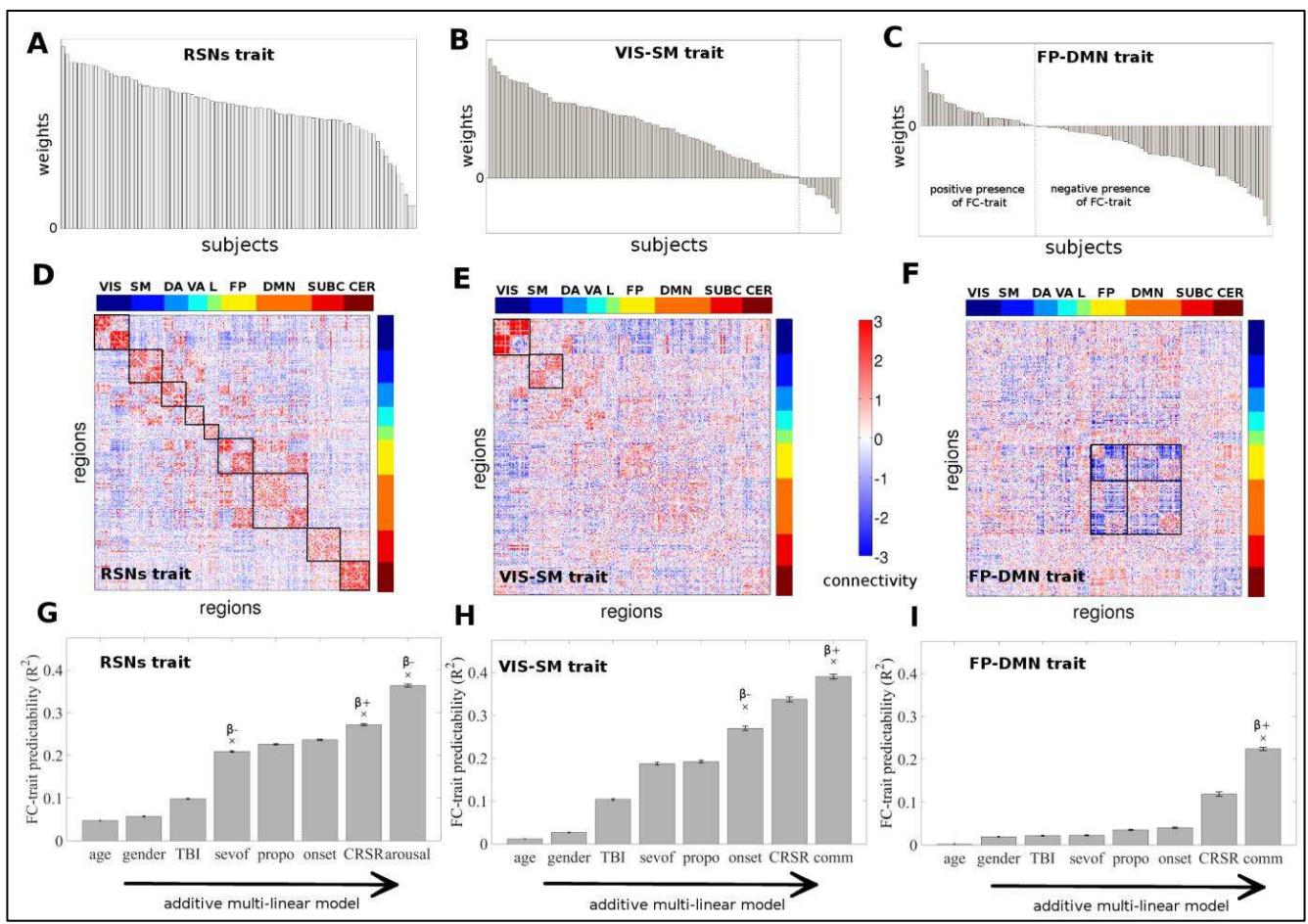

**Figure 3 Mapping of the three main functional traits and their predictability by consciousness features**. **A-C) Quantified presence of each FC-trait on each individual functional connectome.** Subject weights are sorted from greater to smaller on each FC-trait. **D-F) Visualization of the three FC-traits associated to consciousness features.** The brain regions are ordered according to Yeo's (Yeo et al., 2011) functional *RSN*s: Visual (V), Somato-Motor (SM), Dorsal Attention (DA), Ventral Attention (VA), Limbic system (L), Fronto-Parietal (FP), Default Mode Network (DMN), and for completeness, also subcortical regions (SUBC) and cerebellum

(CER). **G-I) Bar-plots of the FC-traits predictability.** They show additive multi-linear regression models with predictors sequentially introduced in the following order: age, gender, trauma, sedative sevof (i.e., sevoflurane), sedative propo (i.e., propofol), inverse of the time since onset, Coma Recovery Scale - Revised (CRS-R) total scores and the CRS-R arousal (for the *RSN*s trait) or communication (for the VIS-SM and FP-DMN traits) subscore. Error bars show the standard error across the 100 ICA runs. Crosses on the top of a bar indicate that the inclusion of the correspondent predictor significantly increased the predictability of the model. The sign of the beta coefficient associated to each significant variable is shown below each asterisk, indicating whether there is a negative or positive trend with respect to the weights of the FC traits.

The positive sign of the beta coefficient associated to the communication subscore indicates that a subject with higher communication subscore has higher contribution or presence of the *VIS-SM* trait in his/her functional connectome (Figure 3B). Interestingly, when adding time since onset (quantified here as the inverse of the days since the insult, see Methods), the explained variance of the model significantly increased. The negative sign of the associated beta coefficients for the onset indicates a negative slope in the fit with the FC individual weights. That is, the more recent the insult, the lower the prominence of the *VIS-SM* trait on the individual FC of the patient.

The trait shown in Figure 3F mainly captures modifications in the connectivity between DMN and fronto-parietal networks (hence denominated *FP-DMN* trait). Interestingly, the *FP-DMN* trait is related to the CRS-R communication subscore, even when the sum of scores is already added to the multi-linear model (Figure 3I). The positive sign of the beta coefficient associated to this predictor indicates that a subject with higher CRS-R sum of scores (communication subscore) has higher presence of the *FP-DMN* trait on his functional connectome. Notably, as one goes lower in the levels of consciousness, the contribution of the *FP-DMN* trait on the FC of a subject changes sign (see the sorted individual weights associated to *FP-DMN* trait, Figure 3C, 3F), with this sign change also evident in a few subjects for the *VIS-SM* trait (Figure 3B, 3E).

Notably, adding the number of censored volumes as an additional predictor was not significant for any of the three multi-linear models discussed above (RSN-trait p=0.65, VIS-SM trait p=0.43, FP-DMN trait p=0.61). Finally, in order to test the appropriateness of the multi-linear models, scatter plots of actual vs predicted responses and of standardized residuals vs predicted responses were also evaluated (see Figure S4 in the Supplementary information).

The *RSNs* trait was mostly characterized by the underlying *RSN*s. We further characterized *VIS-SM* and *FP-DMN* traits by identifying the regions with a higher functional strength. The strength of participation of each brain region to the two FC-traits was measured as its absolute weighted degree (i.e. computed as the sum over columns of the absolute value of the FC-trait). The higher strength indicates more influential role of a brain region to the FC-trait, and hence to the disruption of the level of consciousness. *VIS-SM* trait mainly involves visual areas in the occipital lobe, whereas both medial and lateral fronto-parietal areas are predominant in the *FP-DMN* trait (Figure 4).

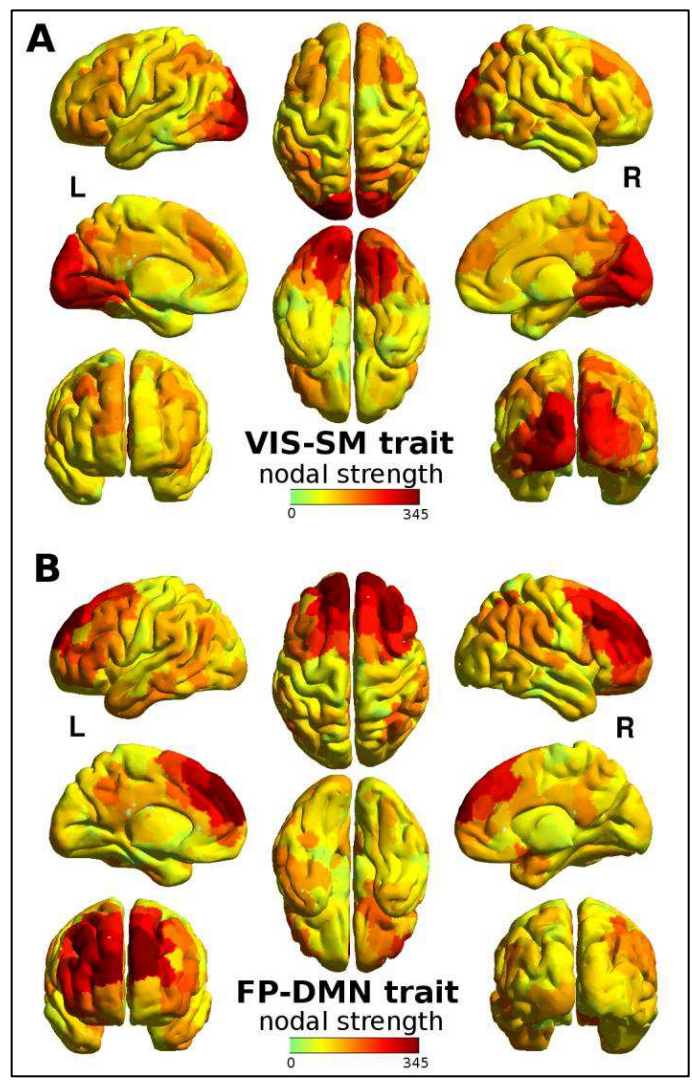

**Figure 4. The strength per region computed as absolute sum of component weights allows an assessment of the overall centrality of each region** for **A)** *VIS-SM* trait and **B)** *FP-DMN* trait. Note how *VIS-SM* trait mainly involves vision-related areas in the occipital lobe, whereas both medial and lateral fronto-parietal dominate in the *FP-DMN* trait.

Further analyses were performed on *VIS-SM* and *FP-DMN* traits to assess the presence of communities (Figure 5) by using consensus clustering (Lancichinetti and Fortunato, 2012) over 100 modularity solutions computed using the Louvain algorithm (Blondel et al., 2008) with quality function Q extended to signed networks (Mucha et al., 2010) (see Methods). Note that the obtained modular configuration gives insights on the data-driven organization of the functional cores common to the whole cohort. When going back to the individual space, the multiplication by the subject's weight may preserve or change this core modular organization depending on the sign (i.e. it changes the FC-trait signed network, see Figure 3). Hence, performing consensus clustering on the FC-traits allowed us to track the "normal conscious" (positive weights) configuration that gets disrupted towards "lower altered" (negative weights) levels of consciousness (Fig. 3A, 3B, 3C).

We looked at the prominence of each community by averaging the correspondent connectivity values within each module. Interactions between every two communities in the FC-traits were then evaluated by averaging the connectivity values connecting them, hence providing a representation of the "coupling" between communities.

For both traits, the highest modularity was associated to partitions consisting of three modules. In line with results of Figure 4, the most influential module (the highest within-module average) for the *VIS-SM* trait appears to be the one comprising the occipital cortex and higher order visual areas. This module is, notably, strongly decoupled from the DMN module (highest between-modules negative connectivity, Figure 5B), suggesting that in a healthy brain these two modules are negatively correlated. This modular configuration is then altered depending on the levels of consciousness.

The modular organization of *FP-DMN* trait revealed a substantial division of the brain in two hemispheres. The between-modules average weight shows that the most "antagonistic" communities encompass the two different hemispheres (Figure 5D), indicating that in normal consciousness the hemispheres are also anti-correlated. This "decoupling" or negative inter-module connectivity might change (i.e. it turns to positive, Figure 3C, 3F) following loss of consciousness.

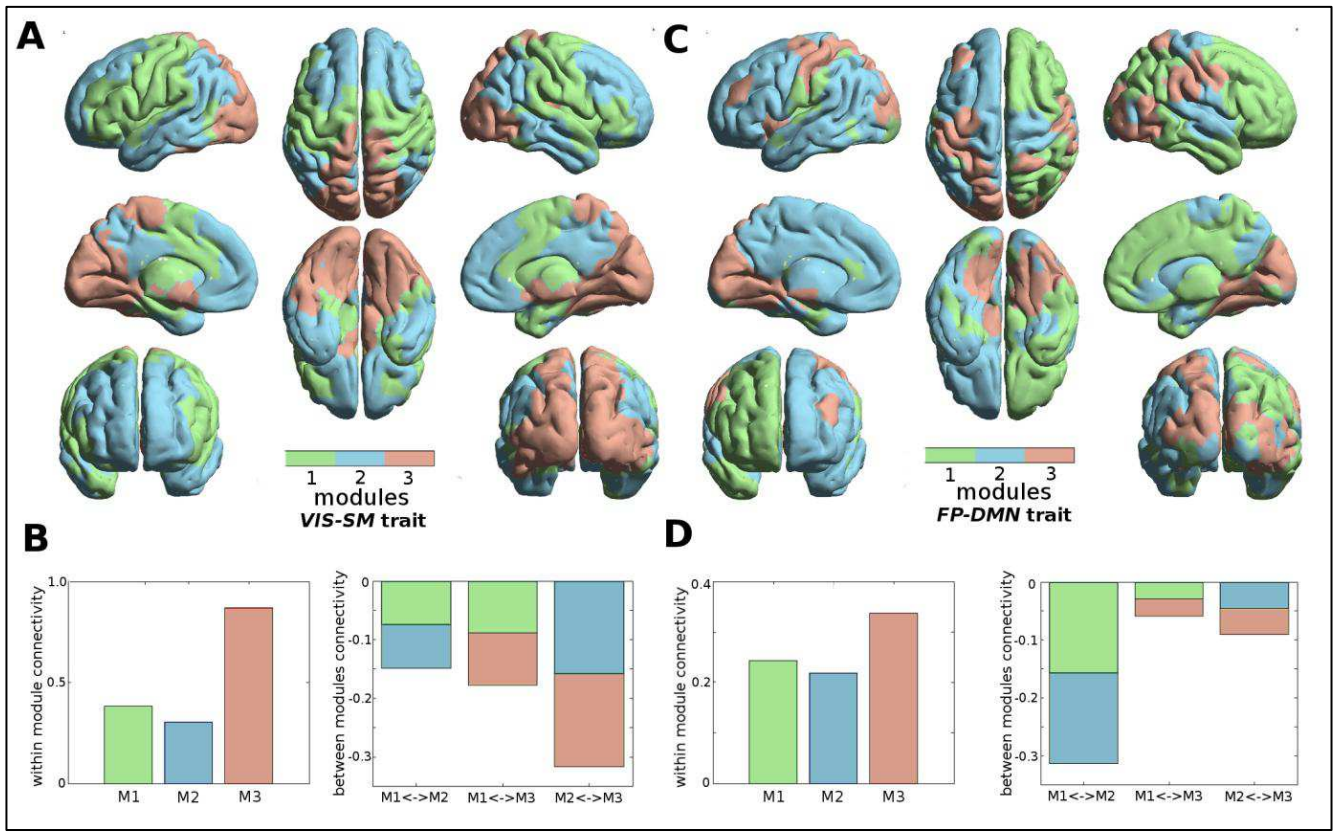

**Figure 5. A) Brain rendering of the modules obtained for the *VIS-SM* trait** (see Materials and Methods). Each color represents region's membership in a module. **B)** Left: bar plot of the average weight within each module in *VIS-SM* trait. Right: bar plot of the average between-module weight for the *VIS-SM* trait. **C) Brain rendering of the modules obtained for the *FP-DMN* trait. D)** Left: bar plot of the average weight within each module in *FP-DMN* trait. Right: bar plot of the average between-module weight for the *FP-DMN* trait.

## Discussion

In this work we applied a novel data-based methodology, *connICA*, to the field of Brain Connectomics. Our approach is based on extracting independent connectivity traits from a set of individual functional connectomes to extract and map robust independent mechanisms or processes that explain the FC patterns of an entire cohort of subjects, without setting any *a priori* stratification into groups. We used the *connICA* framework to assess task-free FC in 88 subjects with different levels of consciousness: 31 conscious controls and 57 severely damaged patients (2 coma, 17 UWS, 20 MCS, 143 EMCS, 4 LIS) of different etiology and duration, 31 of whom were acquired while receiving sedative drugs to control for head movement artifacts. We investigated the functional connectivity traits underlining specific sensorimotor/cognitive capacities related to consciousness.

We showed how these traits separate the FC data into network subsystems with significant associations to levels of consciousness. Notably, this methodology allowed us to map and match the most meaningful functional traits to consciousness-related predictors taken at the patient's bedside. This approach established the link between the alteration of levels of consciousness and the connectivity core associated to it.

The *connICA* framework provides a multiplex data-driven way to extract and compact (dimensionality reduction) the most meaningful multivariate information contained in the functional connectomes in a relatively small set of connectivity traits. In this work we showed how the modification of levels of consciousness is associated to specific connectivity disruptions using as reference seven widely accepted *RSN*s (i.e., visual, somatomotor, dorsal attention, ventral attention, limbic system, fronto-parietal, default mode network (Yeo et al., 2011), and for completeness, also subcortical regions (SUBC) and cerebellum (CER), see insert in Figure 2B).

One additional advantage of this approach is that the dimensionality of the output is significantly reduced, both in the number of the robust components extracted with respect to the initial population size (in the study analyzed here, 5 FC-traits starting from 88 FC matrices) and in the number of variables to be encoded in the multi-linear models, hence notably decreasing number of multiple comparisons. As opposed to univariate approaches mapping up to *n(n-1)/2* functional connections and their subsequent multi-linear models (*n* being the number of brain regions), the multi-layered output of *connICA* results in a small subset of robust FC-traits (by definition, a subset smaller or equal to the number of components set). This dimensionality reduction does not compromise but rather considerably facilitates the interpretability of the results, by compressing the individual variability into the most meaningful independent functional cores. It is noteworthy that most if not all the traits would have been missed with a standard group-average analysis of the functional connectomes.

By using *connICA*, we extracted three independent functional connectivity traits linked to cognitive features of levels of consciousness. Below is a characterization of each FC-trait and its association to different aspects of consciousness.

The *RSNs* trait (Figure 3D) is also the one which explains the most of the FC variance (Figure 2A) and is the closest to the Yeo's *RSN*s organization (Yeo et al., 2011). It seems mainly associated to a global reduction of the functional connectivity within each of the RSNs, and relates to three significant predictors. It was significantly associated to the effect of the pathology (i.e. CRS-R sum of scores and arousal) and to effect of the sedation, specifically to the sevoflurane sedative. Notably, this is in line with several studies reporting the effect of sevoflurane mainly on the functional connectivity of sensory and motor cortices (Deshpande et al., 2010; Hudetz, 2012; Peltier et al., 2005), even at light dosage (i.e. 1-2% concentration, (Peltier et al., 2005), as well as attentional networks (Palanca et al., 2015), frontal and thalamo-cortical networks (Ranft et al., 2016). On the other hand, propofol light sedation at concentrations around 1-2 μg/mL seems to affect mainly the connectivity of fronto-parietal and posterior cingulate cortex (Stamatakis et al., 2010), although in a limited manner as compared to deep sedation (Amico et al., 2014; Boveroux et al., 2010). This may explain why propofol did not show significant associations to any FC-trait in the additive multi-linear models.

The *RSNs-trait* also captures the functional connectivity core organization of a functioning human brain, which gets disrupted during pathological loss of consciousness as evidenced by its strong correlation with the CRS-R sum of scores. This finding is in an agreement with a vast literature showing widespread connectivity breakdowns in several *RSN*s of DOC patients (Boly et al., 2012; Demertzi et al., 2015; Demertzi et al., 2014; Di Perri et al., 2014; Heine et al., 2012; Soddu et al., 2012; Vanhaudenhuyse et al., 2010). One can hypothesize that this general disruption of functional connectivity might be connected and/or partially driven by the widespread underlying structural damage present in DOC and/or TBI patients (Fernández Espejo et al., 2012; Kraus et al., 2007; Perlbarg et al., 2009; Sidaros et al., 2008). Arousal was negatively associated with the *RSNs trait* after controlling for the sum of CRS-R scores as well as other covariates. The assessment of arousal might easily lead to false positive evaluations. For this finding in particular, it is not possible to rule out that the estimation in the level of arousal of a DOC patient may greatly fluctuate and t can thus influence the frequency and complexity of neurobehavioral responses (Schnakers et al., 2008; Seel et al., 2010).

The *VIS-SM* trait seems also associated with the pathology (i.e. time since onset) (Figure 3H). It shows a more prominent disruption of the occipital and sensorimotor areas as the level of consciousness decreases (Figure 3E), and it also correlates with functional communication (Figure 3H). Interestingly, the modularity analysis suggests that visual areas and DMN are anti-correlated in normal wakefulness (Figure 4A, 4B), stressing the importance of the interaction between the so called sensory "slave" regions (Crick and Koch, 1995) and higher order cognitive regions as the DMN, for consciousness and functional communication (Koch et al., 2016). This corroborates the hypothesis that loss of consciousness might correlate with the disruption of primary sensory areas and higher-order associative cortices, which are thought to be required for conscious perception (i.e. global workspace, (Dehaene and Changeux, 2011; Demertzi et al., 2015)).

However, the recovery of this connectivity pattern does not necessarily imply the restoration of levels of consciousness. Another independent functional trait appears to be linked to behavioral assessment of levels of consciousness, particularly to both the CRS-R total score and the communication subscore (*FP-DMN* trait, Figure 3F).

The *FP-DMN* trait captures changes in the anti-correlation between the FP-DMN networks. Notably, as one goes towards the deepest unconsciousness, the FP-DMN anti-correlation decreases, until the point where it "flips" to positive correlation, (see Figure 3C, 3F). This is in line with previous studies showing decreasing anti-correlation in anesthesia (Amico et al., 2014; Boveroux et al., 2010), sleep (Sämann et al., 2011) and UWS patients (Boly et al., 2009). Particularly, a recent study (Di Perri et al., 2016) showed that negative connectivity between DMN and FP networks was significantly different between patients and healthy controls. Indeed, UWS and MCS patients showed a pathological positive connectivity between these two networks, whereas patients who emerged from MCS and recovered a level of consciousness sufficient for functional communication and/or object use, exhibited partial preserved between-network negative connectivity (Di Perri et al., 2016). In this respect, the fact that the *FP-DMN* trait is strongly correlated to the communication subscore corroborates the idea that recovery of the FP-DMN between-network negative connectivity is prerequisite in order to regain functional communication.

Notably, the modularity analysis on *FP-DMN* trait reveals that the *decoupling* between the two hemispheres (Figure 5C) represents a "healthy" way of communication between left and right brain hemispheres. The anti-correlation between hemispheres tends to disappear (i.e. trends toward zero or even becomes positive correlation, see the individual weights of *FP-DMN* trait in Figure 3C) as levels of consciousness decrease.

Indeed, there is evidence suggesting that coordination between the two hemispheres is essential for a correct communication between them (Gazzaniga, 2005). It has been reported that transection of corpus callosum in refractory epileptic patients (i.e. split brain patients) caused each hemisphere to have its own separate perception, concepts, and impulses to act (Gazzaniga, 2014). The conscious abilities of the two hemispheres are strongly differentiated in specialized cognitive modules (Marinsek et al., 2014), modulated by the thalamo-cortical system (subcortical regions are also split in left and right modules in *FP-DMN* trait, see Figure 5C). In this study we show that the interaction between specialized modules, as the VIS-SM interaction with DMN or the FP-DMN between-network negative connectivity, is crucial for the emergence of consciousness. Perhaps this laterality enhances the complexity of ongoing brain processes and facilitates demanding cognitive processes such as consciousness of the self and the surrounding.

Taken together, these findings suggest that the connectivity core which differentiates across levels of consciousness is a combination of positive and negative interactions between functional sub-networks. This evidence stresses the importance of a whole-brain network modulation between coherent and non-coherent functional states. The disruption of the equilibrium between these two might lead to changes in levels of consciousness and, ultimately, to reduced levels of consciousness.

In fact, the *connICA* results presented in this study depict a very challenging reality. Within the set of individual functional connectomes analyzed here, there is not just one but at least three independent mechanisms, namely FC-traits, whose predictability by consciousness related features is present but different for each one, and hence is most likely capturing different phenomena or mechanisms. The first *RSNs* trait predicted by the CRS-R sum of scores, isolates the functional connectivity blocks of typical *RSN*s present in a human brain (Figure 3D). The second *VIS-SM* trait, with predominant influence of visual and sensory regions, relates disruption of sensory networks to the CRS-R functional communication subscore (Figure 3E). The third *FP-DMN* trait, significantly associated to CRS-R sum of scores and communication, stresses the key role of the negative connectivity between FP and DMN networks (Figure 3F) and inter-hemispheric communication (Figure 5C,D) in the alteration of levels of consciousness.

It is worth mentioning here that the traits found by *connICA* and the sensitivity of those to demographical and cognitive features is highly dependent on the population analyzed. Indeed, when considering the *RSN*s trait, demographics such as age appear to have a strong fingerprint on it when looking only to the healthy cohort (without considering DOC patients, see supplementary Figure S5). This suggests that age has a fingerprint in FC-traits obtained from a healthy population, but its relative effect is blurred when assessing subjects at different levels of consciousness (metaphorically being the latter "the elephant in the room").

The study presented here adds to recent studies from Iraji et al. (Iraji et al., 2016) assessing ICA components of voxel-based functional connectivity, and from Misic et al. (Misic et al., 2016), where levels of integration of joint structural-functional connectivity patterns are assessed from sets of individual connectomes by means of a single-value decomposition (Misic et al., 2016). Together with the methodology presented here, these recent efforts suggest that the area of Brain Connectomics is evolving into new data-driven ways of analyzing connectivity data at different spatial scales without stratifying subjects into *a priori* groups and hence, also without performing group-averages of individual connectivity matrices.

Our study has several limitations. The optimal size of the cohort for the extraction of the *connICA* components needs to be further investigated. Similarly, the best choice of the starting number of ICA components (here set to 15) and the threshold for the final selection of the most frequent components over multiple ICA runs (here set to 75%) need to be characterized in more detail. In this work we used the Shen brain parcellation (Shen et al., 2013) because of the uniformity of the size of brain regions and its functional data-driven approach. We also used the well-assessed *RSN*s decomposition provided by Yeo as obtained in a large cohort (n=1000) of healthy volunteers (Yeo et al., 2011). However, other parcellations (Desikan et al., 2006; Gordon et al., 2016) or finer decompositions (Demertzi et al., 2015; Demertzi et al., 2014) might be beneficial in the *connICA* framework, depending on the research problem at hand and the desired level of spatial resolution.

Future work can be extended to the use of *connICA* for structural connectivity patterns, hence identifying SC-traits within a population of subjects. This approach is not limited to assessing consciousness, but it has the potential of studying other progressive diseases and disorders, drug-induced effects, and also differences based on aging or gender. We have here addressed the effect of the sedation as a binary confound (see Materials and Methods). An interesting future avenue would be to apply *connICA* for disentangling differences between FC-traits at different concentrations of the anesthetic agent at hand, e.g. in a population of healthy subjects.

When associating traits with cognitive/clinical features, multi-linear models employed here can be expanded by allowing for non-linear terms and interactions, which could capture more complex associations between connectivity patterns and cognition.

In conclusion, we here proposed a novel data-driven approach, *connICA*, to extract the most influential connectivity patterns in the alteration of levels of consciousness. Our results shed light on isolating key functional core changes involved in the degradation of conscious states and establish links between isolated clinical/cognitive features and specific FC-traits.

## Acknowledgements

We thank Marie-Aurelie Bruno, Athena Demertzi and Audrey Vanhaudenhuyse for help in acquiring the data. We thank Prof. Jaroslaw Harezlak and Prof. Thomas Talavage for useful comments. This research was supported by the This research was supported by the Belgian Funds for Scientific Research (FRS), European Commission, James McDonnell Foundation, European Space Agency, Belgian Science Policy (CEREBNET, BELSPO), Wallonia-Brussels Federation Concerted Research Action, Mind Science Foundation, Public Utility Foundation "Université Européenne du Travail" and "Fondazione Europea di Ricerca Biomedica", University and University Hospital of Liège. LH is a research fellow and SL a research director at FNRS. JG was supported by the National Institute of Health (1R01 MH108467-01).

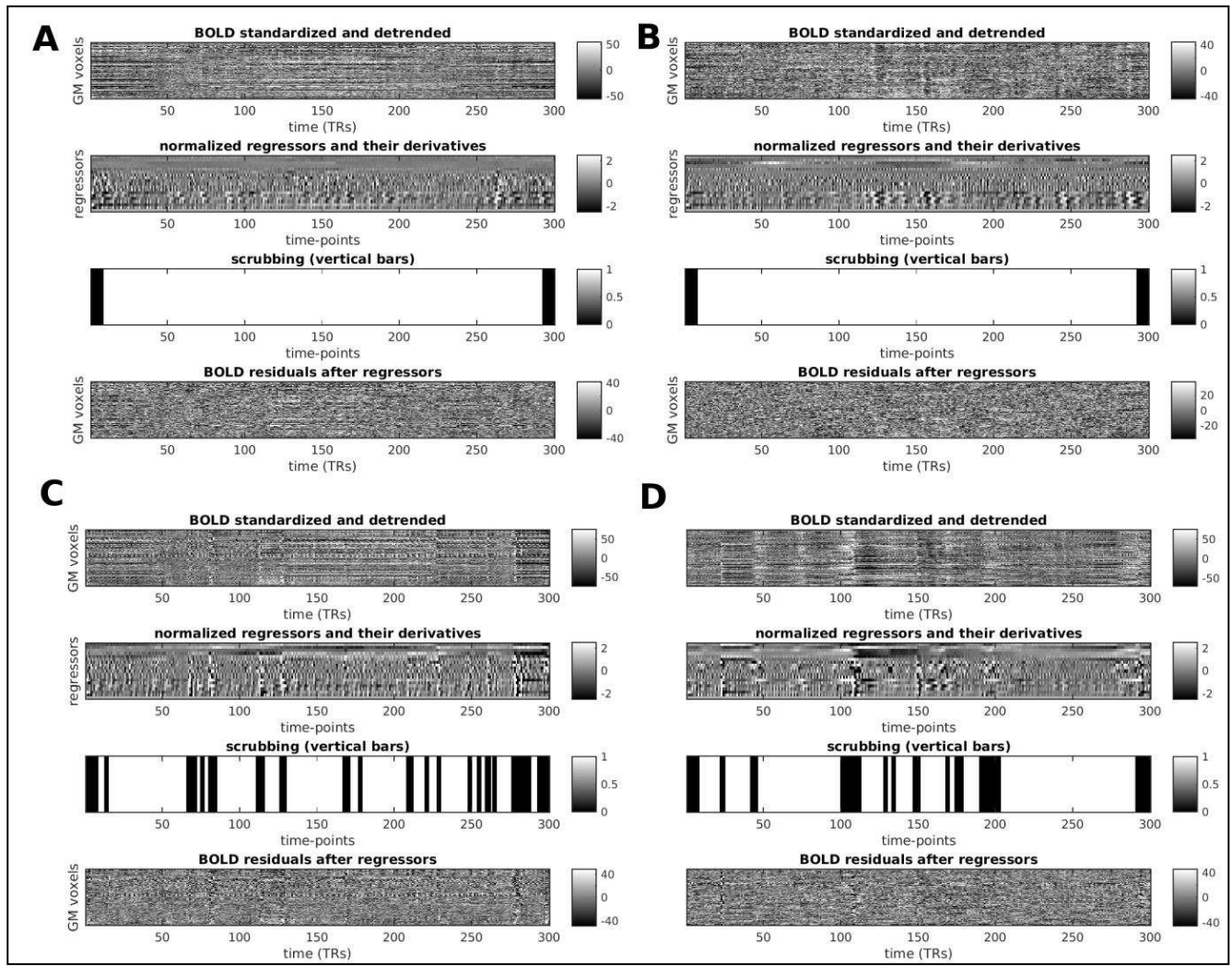

**Figure S1: Illustration of the major fMRI preprocessing steps** described in the Materials and Methods section of the task-free session for four individual subjects (A-D). For each subject, the 4 plots from top to bottom: 1) The fMRI time courses in all GM voxels after slice timing and motion correction, normalization to mode 1000, demeaning and detrending; 2) The 18 motion and physiological noise regressors; [x, y , z, pitch, yaw, roll], the tissue mean signal of whole-brain, WM and CSF and their corresponding nine derivatives (backwards difference); 3) Visual representation of the scrubbing procedure using the Frame Displacement (FD), DVARS and SD metrics to drop (censor) BOLD volumes with head motion (indicated by the dark vertical bars) from the computation of the pairwise correlations. Note that, as explained in Materials and Methods, first and last 7 volumes in each session were always excluded; 4) Residuals of the BOLD time courses of GM voxels after regressing out the 18 regressors. Subjects A and B had no censored volumes, with C and D having 14% and 11% censored volumes, respectively.

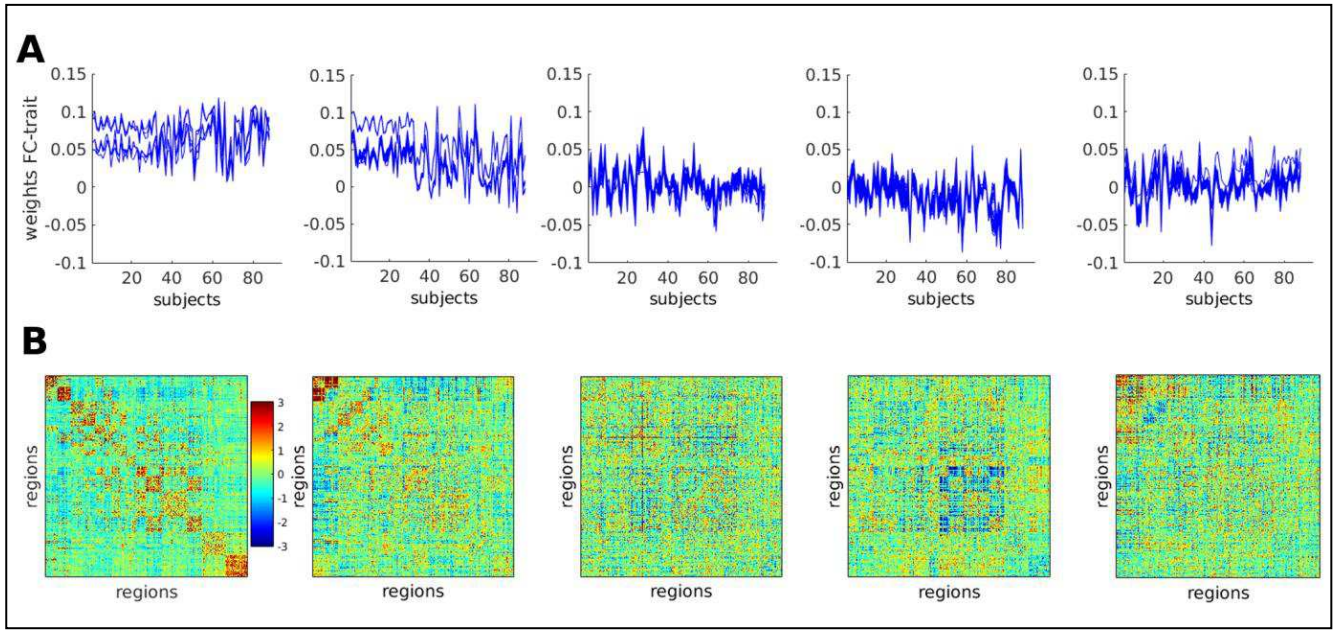

**Figure S2: A)** Plot of the correspondent weights per subjects for each single *connICA* run. **B)** Plot of the five robust connectivity traits extracted using *connICA* based on 100 runs. Only three of the FC-traits were associated with cognitive scores linked to levels of consciousness.

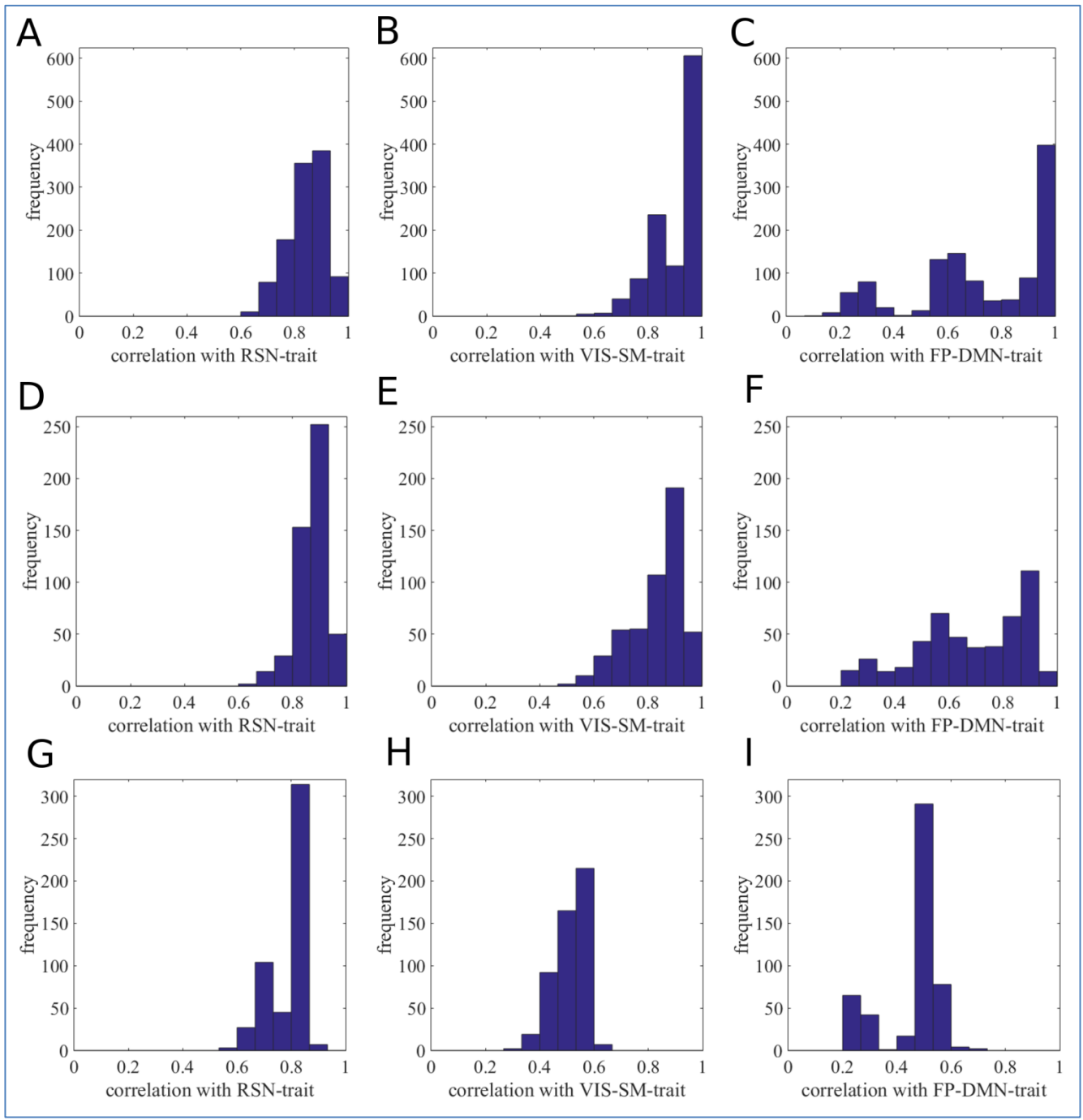

**Figure S3: Robustness analysis of FC-traits. A-B-C) Robustness of FC-traits to number of components.** Histograms of the correlations between each FC-trait and its best matching component for each of the 1,100 *connICA* runs performed. Those runs consisted of running *connICA* while varying the number of components between 10 and 20 (100 runs for each). This test was meant to check whether the choice on the number of components would "drive" the computation of one of the 3 FC-traits found. **D-E-F) Robustness of FC-traits to data sampling.** Histograms show the correlations between the components that best match each of the presented 3 FC-traits. Here, the number of components is fixed to 15, but random subsamples of the 80% of the initial dataset are included for every run (500 runs) to extract the components. This allowed us to evaluate whether the generation of the observed FC-traits was "driven" by any subset of subjects. **G-H-I) Robustness of FC-traits to exclusion of sedated subjects.** Histograms show the correlations between the components that best match each of the presented 3 FC-traits. Here, the number of component is fixed to 15, but sedated patients were excluded from the dataset used to extract the components (500 runs). This allowed us to evaluate if the generation of the observed FC-traits was "driven" by the sedated patients.

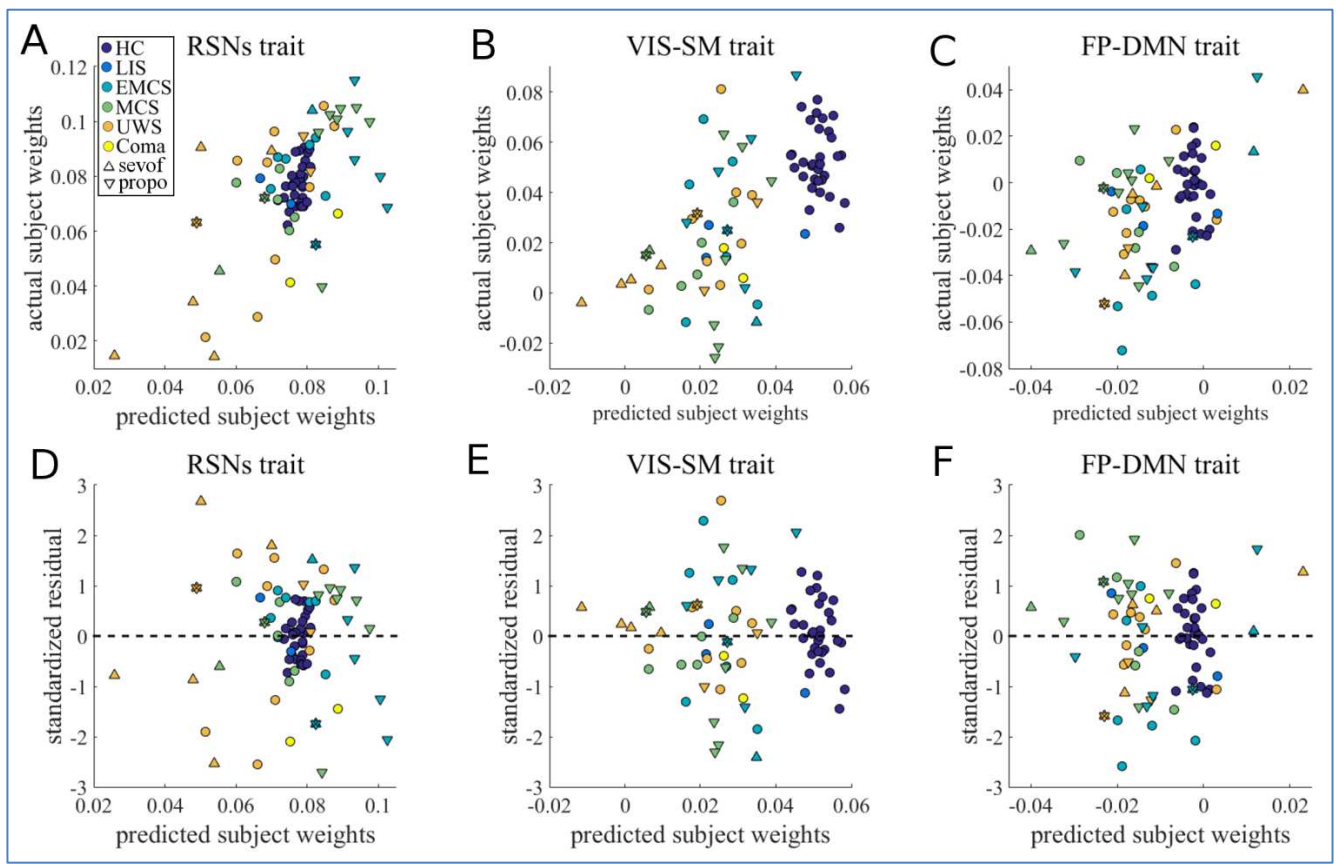

**Figure S4: Evaluation of the multi-linear models. A-B-C) Predicted vs actual values.** Scatter plots of the subject weights (Y-axis) associated to each of the 3 extracted FC-traits (i.e., RSNs, VIS-SM and FP-DMN) versus the subject weights predicted by the final multi-linear regression models when including all 8 predictors (see Material and Methods and Figure 3 for details and description of predictors). Colors denote diagnosis as follows: dark blue refers to Healthy Controls (HC); blue to Locked-in Syndrome patients (LIS); light blue to patients emerging for minimally conscious state (EMCS); green to minimally conscious state patients (MCS); orange to unresponsive wakefulness syndrome patients (UWS); yellow to coma patients (Coma). Shape denotes sedatives administered to subjects: circles indicate non-sedated, upward-pointing triangles indicate sevoflurane, downward-pointing triangles indicate propofol, and hence stars indicate both sevoflurane and propofol. Note how the linear trend between actual-predicted values is not driven by any specific group or a sedative, providing evidence on the reliability of the employed model. **D-E-F) Predicted vs residuals.** Scatter plots of the standardized residuals (Y-axis) versus the predicted subject weights associated to each of the 3 extracted FC-traits (i.e., RSNs, VIS-SM and FP-DMN) for the multi-linear model described above (see also Material and Methods for details). The residuals are symmetrically distributed, tending to cluster around 0, and within 3 standard deviations of zero.

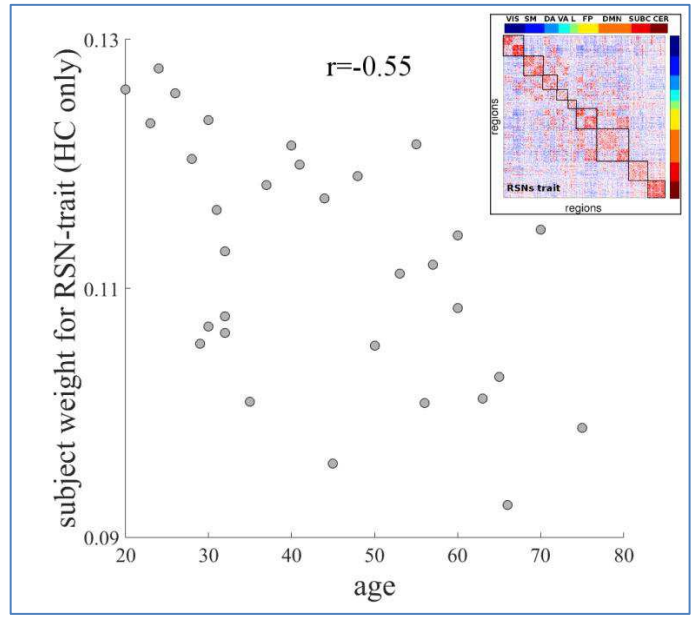

**Figure S5. Scatter plot of age vs RSNs-trait subject weights in HC group.** When looking at HC subjects only (N=31), we found a significant negative correlation (r=-0.55, p<0.05) between age and the subject-weight for RSNs-trait (top right insert). In other words, the contribution of the RSNs-trait in the resting-state FC is reduced in older subjects.

| Trait/Predictor | Age | Gender | TBI | Sevoflurane | Propofol | Onset | CRS-R | Subscore |
|---|---|---|---|---|---|---|---|---|
| **RSNs** | N.S. | N.S. | N.S. | -0.0079 | N.S. | N.S. | 0.0156 | -0.0143 |
| **VIS-SM** | N.S. | N.S. | N.S. | N.S. | N.S. | -0.0074 | N.S. | 0.0188 |
| **FP-DMN** | N.S. | N.S. | N.S. | N.S. | N.S. | N.S. | N.S. | 0.0206 |

**Table S1.** Standardized beta coefficients of the final models for each FC trait and including all predictors (see Materials and Methods for details on each predictor). N.S. denotes non-significant values. Statistical significance set at p < 0.05.